\documentclass[english,12pt,twoside]{article}
\usepackage{amssymb}
\usepackage{amsmath}
\usepackage[pdftex]{graphicx}
\usepackage{pst-node}
\begin{document}

\def\meter{\mbox{$\frown\hspace{-.9em}{\lower-.4ex\hbox{$_\nearrow$}}$}}

\date{}
\title{Transitivity of an entangled choice}
\author{Marcin Makowski\footnote{makowski.m@gmail.com}, Edward W. Piotrowski\\[1ex]\small
Institute of Mathematics, University of Bia\l ystok,\\\small
Akademicka 2, PL-15424, Bia{\l}ystok,
Poland}
\maketitle\begin{abstract}
 We describe a quantum model of simple choice game (constructed upon entangled state of two qubits), which involves the fundamental problem of transitive -- intransitive preferences.  We compare attainability of optimal intransitive strategies in both classical and quantum models with the use of geometrical interpretation. 
\end{abstract}
\section{Introduction}\noindent
This paper is a continuation of our  research on intransitive orders in quantum game models, which was reported  in \cite{r2,rMM}. 
We will consider some possible modifications of the previously analysed game \cite{r1}, which will eliminate certain significant limitations resulting from the original model's assumptions. The selecting player  (a cat) will now have a wider range of options during the decision--making process. This will allow the cat to choose an optimal strategy, regardless of the offering player's (Nature) move (this element was absent in the previous model). Let us analyse in what way the absence of the above--mentioned limitation may influence the occurrence of optimal intransitive strategies. This is a continuation of previous analyses, and the game conditions will now be much more advantageous for the selecting player. What influence does this have  on the significance of intransitive strategies? We will answer this question by offering a geometrical interpretation of the game, which will allow us to illustrate the attainability of different optimal strategies. We shall also introduce a simple and natural representation of the quantum model by taking advantage of the entangled state of two qubits, assuming that one of them corresponds with Nature's moves and the second describes the cat's strategy (preference). This is a novel approach to the quantitative analysis of various types of strategies.     
\vspace{0.2cm}\\
Moreover, we will have a closer look at how alterations of game conditions and modelling methods influence the occurrence of intransitive orders, which are deemed a mysterious and surprising phenomenon of nature.  These issues still await a thorough analysis within the quantum theory which has already proved to be useful for describing and explaining reality. Despite numerous research studies on usefulness of quantum formalism in constructing  games and algorithms \cite{r3,r4,r5,r6,r7,r8,r9,r11,r12}, the issue of intransitive orders is rarely subject to analysis. For the time being, only a few papers aimed at analysing the problem have been written \cite{tt,tt1}. Considering the importance of intransitivity for classical games and decision theory (see Sec. 2), one may assume that it will be  important for research on characteristics of games analysed within quantum information theory. Recognising their importance may be achieved only in the progress of further research.  
\section{Intransitivity}\noindent
Any relation $\succ$  between the elements of a certain set is called \emph{transitive} if $A\succ C$ results from the fact that $A\succ B$ and $B\succ C$ for any three elements  $A$, $B$, $C$. If this condition is not fulfilled then the relation will be called \emph{intransitive} (not transitive).
\vspace{0.2cm}\\
There exists a prevailing notion that people who make decisions relying on rational reason, should make decisions in determined and linear order. This axiom is the fundamental element of classical theories of individual and collective choices \cite{r22}. One of the main arguments put forward by many experts, proving the irrationality of preferences which violate transitivity, is the so -- called ``money pump'' (see \cite{r13}). 
This notion can be intuitively comprehended and arises from the fact that transitivity determines our inference processes (a type of reasoning called ,,transitive inference'': $A\succ B$ and $B\succ C$ imply $A\succ C$) since human are 4 -- 5 years old \cite{r33}. Acceptance of this rule enables children to shape such skills as measuring, arranging elements and deductive thinking. This characteristic is considered by biologists to be an essential element of logical thinking, therefore most of researches devoted to animal ways of thinking focus on the rule of transitive inference \cite{r33}. 
Despite the fact that intransitivity appears to be contrary to our intuition, nature provides many examples of intransitive orders. Rivalry between species may be intransitive. For example, in the case of fungi, Phallus impudicus replaced Megacollybia platyphylla, M. platyphylla
replaced Psathyrella hydrophilum, but P. hydrophilum replaced P. impudicus \cite{r15}. Also bees make
intransitive choices between flowers \cite{r16}. In chemistry, one of the best known examples of intransitive order is the so--called Belousov--Zhabotinsky reaction, in which different colors of liquid sequentially replace one another again and again. Belousov had enormous problems with the publication of his discoveries for many years, with many chemists thinking that such a reaction is impossible \cite{Biel}.

Analyses of intransitive orders became a significant element of researches concerning the decision theory. Kenneth Arrows' theorem \cite{r24} is the most spectacular achievement in this area. He proved that an election procedure which would perfectly fulfil basic postulates of democracy does not exist. Arrow was referring to Condorcet's work published in 1785, where he examined voting paradox indicating something astonishing: collective preferences can be may be intransitive even if the preferences of individual voters are not.

In psychology, which attempts to describe and explain the decision making process, the special interest is focused on a broadly understood relation of domination (superiority, preference). Does the fact that  \upshape{A} predominates over  \upshape{B} and \upshape{B} predominates over \upshape{C}, implies that \upshape{A} predominates over \upshape{C}? The answer is not as obvious as it may appear, and because the question is so general, the answer depends on a particular situation (and how do we interpret predomination). This issue is widely discussed. Some scientists argue that examples provided by others \cite{Chan} do not really prove intransitivity of the generally interpreted relation of domination and claim that transitivity of preferences for available options is part and parcel of rational decision making process \cite{tull}. Others think that theory of  rational decision -- making does not necessarily have to be based upon the ``axiom of transitivity'' \cite{r22,Bau,irra}. The problem consists in determining which relation is intransitive. This call for establishing a general rule  allowing to solve this problem \cite{Rob}.            
 \vspace{0.2cm}\\
Significance of intransitive orders for explanation of rational decision -- making process and for description of natural phenomena, fully justifies the need for analysing this subject within the quantum game theory. This new approach of modelling the decision algorithms (in relation to the quantum information) may lead to many interesting conclusions and observations.
\section{Description of the model}\noindent
We will reconsider the game described and examined in  \cite{r1}, where a cat is a character. Let us assume that a cat is offered three types of food (no. 0, no. 1 and no. 2), every time in pairs of two types, whereas the food portions are equally attractive regarding the calories, and each one has some unique components that are necessary for the cat’s good health. The cat cannot consume both offered types of food at the same moment and it will never refrain from making the choice.
Let us introduce a random factor to the game (the ``automatic machine'') and let us assume that the offering player (Nature) decides about offering a particular type of food, but the selected type of food is always offered with one of the remaining types of food. The offering player does not favour any of the two available pairs which contain the selected food (the probability of offering is the same -- 1/2). At this stage of the game the player, i.e. Nature, behaves like an ,,automatic machine'' and with equal probability offers two available food pairs which contain the selected type of food. 
\subsection{Mathematical model}\noindent
Let us denote by $q_i$ the frequency of choosing (by Nature) the food of number $i$. It is, in fact, the frequency of occurrence
of two pairs of food which contain food number $i$ (each of two pairs are then offered with probability 1/2). We will denote by $P_i(C_{k}|B_{j})$ the probability of choosing the food of number $k$, when the offered food pair does not contain the food of number $j$ (in case when Nature choose food no. $i$). 
There is the following relation between the frequencies $\omega_k$, k = 0, 1, 2, of appearance of the particular foods in a diet and the conditional probabilities:
\begin{itemize}
	\item food no. 0:
	\newline $\omega_0=P_0(C_{0}|B_{2})\frac{1}{2}q_0+P_0(C_{0}|B_{1})\frac{1}{2}q_0 +P_1(C_{0}|B_{2})\frac{1}{2}q_1+P_2(C_{0}|B_{1})\frac{1}{2}q_2,$
	\item food no. 1:
	\newline $\omega_1=P_0(C_{1}|B_{2})\frac{1}{2}q_0+P_1(C_{1}|B_{2})\frac{1}{2}q_1 +P_1(C_{1}|B_{0})\frac{1}{2}q_1+P_2(C_{1}|B_{0})\frac{1}{2}q_2,$
	\item food no. 2:
	\newline $\omega_2=P_0(C_{2}|B_{1})\frac{1}{2}q_0+P_1(C_{2}|B_{0})\frac{1}{2}q_1 +P_2(C_{2}|B_{1})\frac{1}{2}q_2+P_2(C_{2}|B_{0})\frac{1}{2}q_2.$
\end{itemize}
The most valuable way of choosing the food by cat occurs when:
\begin{equation}
 \label{maximum}
 \omega_0=\omega_1=\omega_2=\frac{1}{3}.
 \end{equation}
 Any twelve conditional probabilities $\{P_i(C_{k}|B_{j})\}$, that for a fixed triple $(q_0, q_1, q_2)$ fulfill (\ref{maximum})
will be called a \textsl{cat’s optimal strategy}.

The system of equations (\ref{maximum}) has the following matrix form:
 \begin{eqnarray}\label{matrix maximal}
B\left( \begin{array}{ccc} q_0 \\ q_1 \\ q_2
\end{array} \right)=
\frac{1}{3}\left( \begin{array}{ccc} 1 \\ 1 \\ 1
\end{array} \right),
\end{eqnarray}
where:
\begin{eqnarray}
B=\left( \begin{array}{ccc} \frac{P_0(C_0|B_2)+P_0(C_0|B_1)}{2} & \frac{P_1(C_0|B_2)}{2} & \frac{P_2(C_0|B_1)}{2} \\ \frac{P_0(C_1|B_2)}{2} & \frac{P_1(C_1|B_2)+P_1(C_1|B_0)}{2} & \frac{P_2(C_1|B_0)}{2} \\ \frac{P_0(C_2|B_1)}{2} & \frac{P_1(C_2|B_0)}{2} & \frac{P_2(C_2|B_1)+P_2(C_2|B_0)}{2}
\end{array} \right).\nonumber
\end{eqnarray}
The solution of equations (\ref{matrix maximal}): 
\begin{eqnarray}\label{odw}
q_0&=&\frac{1}{d}\bigg(2-P_2(C_{0}|B_{1})-2P_2(C_{1}|B_{0})+2P_1(C_{1}|B_{0})\nonumber\\\nonumber &+&3P_1(C_{0}|B_{2})P_2(C_{0}|B_{1})- 3P_1(C_{1}|B_{0})P_2(C_{0}|B_{1})\\\nonumber &+&3P_1(C_{0}|B_{2})P_2(C_{1}|B_{0})-4P_1(C_0|B_2)\bigg),\nonumber\\
q_1&=&\frac{1}{d}\bigg(-2+P_2(C_{0}|B_{1})+2P_2(C_{1}|B_{0})+2P_0(C_{0}|B_{1})\\\nonumber&-&3P_2(C_{1}|B_{0})P_0(C_{0}|B_{2})-3P_2(C_{0}|B_{1})P_0(C_{0}|B_{2})\\\nonumber&-&3P_2(C_{1}|B_{0})P_0(C_{0}|B_{1})+4P_0(C_0|B_2)\bigg),\nonumber\\
q_2&=&1-q_0-q_1,\nonumber
 \end{eqnarray} 
 defines a mapping $A: D_6\rightarrow T_2$\, of the six--dimensional cube $D_6$ conditional probabilities into a two--dimensional simplex (triangle) $T_2$ ($d$ is the determinant of the matrix $B$). The barycentric coordinates of a point of this triangle are interpreted as frequencies $(q_0, q_1, q_2)$. Thus we get relation between the optimal cat’s strategy and frequencies $q_i$ of appearance of individual food. 
\vspace{0.2cm}\\
Different  meaning of  $q_i$ parameters in the described model and in the previous version \cite{r1}, influences particular characteristics distinguishing these models. In the paper \cite{r1} the cat on each stage was offered a pair of food and forced to choose one of them. Each pair was appearing with determined frequency and the cat was aware of this fact. The parameter $q_i$, $i\in \{0,1,2\}$ denoted frequency with which the pair of food without $i$ food was appearing. In a model of such construction, problem of existence of an optimal strategy had significant importance, as with certain limitations of $q_i$ parameters this type of strategy was unattainable (when $q_i>2/3$ for some $i$).                      
In the described model, optimal strategy is available regardless of the frequency  $(q_0, q_1, q_2)$. 
This is a simple implication of the assumption that $q_i$ parameter marks the frequency with which the offering player (Nature) decides to offer food $i$, but (what is important) offers it always (``automatic machine'') in combination with one of the two remaining types of food with  probability $1/2$. This means that even if for certain $i$ and $j$, ($i,j\in \{0,1,2\}$ and $i\neq j$) we will have $q_i=q_j=0$, the cat will still have a chance to choose one of three possible types of food. This is crucial for existence of an optimal strategy.
\vspace{0.2cm}\\
Prior to discussing the quantum game model, let us write inverse  transformation of the mapping defined by (\ref{odw}), and provide an example of optimal strategy.
After introducing parameters  $\alpha$, $\beta$, $\gamma$, $\delta$ it will go as follows:    
$$P_0(C_{0}|B_{2})=\frac{-2 + 3(-\alpha +\gamma+\delta+q_0+q_1)}{3q_0},$$
$$P_0(C_{0}|B_{1})=\frac{1 + 3(-\beta-\gamma-\delta + q_2)}{3q_0},$$
$$P_1(C_{0}|B_{2})=\frac{\alpha}{q_1},$$
$$P_2(C_{0}|B_{1})=\frac{\beta}{q_2},$$
$$P_1(C_{1}|B_{0})=\frac{\gamma}{q_1},$$
$$P_2(C_{1}|B_{0})=\frac{\delta}{q_2}.$$
The properties of probability impose following conditions on parameters $\alpha$, $\beta$, $\gamma$, $\delta$:
\begin{eqnarray}\label{waropt}
&\alpha, \gamma \in [0,q_1], \nonumber\\ &\beta, \delta \in [0,q_2],\nonumber\\
&\frac{2}{3} - q_0 - q_1 \leq - \alpha + \gamma + \delta \leq \frac{2}{3} - q_1,\\ &\frac{1}{3} + q_2- q_0\leq\beta+\gamma+\delta\leq\frac{1}{3} + q_2.\nonumber
\end{eqnarray}
\subsection{Example of an optimal strategy}\noindent
Let us assume that Nature decides to provide given types of food with $q_0=\frac{20}{24}$, $q_1=\frac{1}{24}$, $q_2=\frac{3}{24}$, frequencies. In order to establish an optimal strategy, it will be necessary to determine parameters $\alpha$, $\beta$, $\gamma$, $\delta$ which fulfills conditions (\ref{waropt}), and for the provided $q_m$ frequencies look as follows:  
\begin{eqnarray}
&\alpha, \gamma \in [0,\frac{1}{24}], \nonumber\\ &\beta, \delta \in [0,\frac{3}{24}],\nonumber\\
&-\frac{5}{24} \leq - \alpha + \gamma + \delta \leq \frac{15}{24},\\ &-\frac{9}{24}\leq\beta+\gamma+\delta\leq\frac{11}{24}.\nonumber
\end{eqnarray}
By selecting: $\alpha=\gamma=\frac{1}{48}$, $\beta=\delta=\frac{1}{24}$, we get an optimal strategy:
\begin{eqnarray}\nonumber
P_0(C_{0}|B_{2})& =\frac{3}{10}\,, \qquad&P_0(C_{0}|B_{1})=\frac{17}{40}\,, \\\nonumber
P_1(C_{0}|B_{2})&=\frac{1}{2}\,,\qquad&P_2(C_{0}|B_{1})=\frac{1}{3}\,,\\\nonumber
P_1(C_{1}|B_{0})&=\frac{1}{2}\,, \qquad&P_2(C_{1}|B_{0}) =\frac{1}{3}\,.\nonumber
\end{eqnarray} 
\section{Quantum model}\noindent
Now, we will describe the quantum game model. It is based upon the assumption that Nature and the cat perform actions on the system of two qubits remaining in the EPR state (maximal entanglement):
\begin{equation}\label{spl}
 \frac{1}{\sqrt{2}}
(|0\rangle\negthinspace\,|0\rangle\negthinspace + |1\rangle\negthinspace\,|1\rangle\negthinspace)\,.
\end{equation} 
We decide that the right qubit belongs to the cat and the left one is connected with Nature's move. 
Selection of the given type of food by the offering player (Nature) indicates selection of one out of three mutually unbiased bases in two--dimensional Hilbert space $H_2$\,:
\begin{eqnarray}
 \{|0\rangle\negthinspace\, ,  |1\rangle\negthinspace\, \}:=&\{(1,0)^{T},(0,1)^{T}\}, \nonumber\\ 
 \vspace{6pt}
 \{|0'\rangle\negthinspace\, ,  |1'\rangle\negthinspace\, \}:=&\bigg\{\frac{| 0 \rangle+| 1 \rangle}{\sqrt{2}}, \frac{| 0 \rangle-| 1 \rangle}{\sqrt{2}}\bigg\},\nonumber\\
 \{|0''\rangle\negthinspace\, ,  |1''\rangle\negthinspace\, \}:=&\bigg\{ \frac{| 0 \rangle+i | 1 \rangle}{\sqrt{2}},\frac{| 0 \rangle-i| 1 \rangle}{\sqrt{2}}\bigg\},\nonumber
 \end{eqnarray}
 and measurement of the left qubit in this base. These bases play an important role in universality of quantum market games \cite{r28} (and allowed Wiesner
 to begin research into quantum cryptography \cite{r27}). Vectors marking this base correspond to two possible pairs containing the selected type of food (i.e. if the cat selects type of food no. $0$, it will be offered a pair $(0,1)$ or $(0,2)$ and these pairs are respectively described by vectors of the base corresponding to the selection of type of food no. 0). Vector coordinates (squares of their moduli), represent probabilities with which a given pair containing the selected type of food is offered. The left qubit acts as an ``automatic machine'' separating the food pairs.    
In order to determine its preference for food pairs, the cat acts on the right qubit with the use of tactics parameterized by the matrix (Euler--Rodrigues parametrization): 
\begin{equation}\label{sfera}
H=\left(\begin{array}{cr}x_1+ix_2 & x_3+ix_4 \\ -x_3+ix_4 & x_1-ix_2\,
\end{array}\right):=\left(\begin{array}{cr}a & b \\ c & d\,
\end{array}\right),
\end{equation}
where $(x_1,x_2,x_3,x_4)\in S_3$. 

Subsequently, coordinates of the right qubit, read (measured) in three mutually unbiased bases of $H_2$ space, determine conditional probabilities which are our concern (cf. \cite{r2}).   
Let us assume that measurement of the left qubit in the standard base  $\{|0\rangle\negthinspace\, ,  |1\rangle\negthinspace\, \}$ is connected with offering by Nature food no. 0  (vector $|0\rangle\negthinspace$ represents pair $(0,1)$ and vector  $|1\rangle\negthinspace$ pair $(0,2)$).  Measurement in the standard base of the right qubit indicates the cat's preference for pair $(0,1)$. Roles of the bases and given base vectors in the quantum game model are presented in the Table \ref{tabela222}.  
\begin{table}[h!t]
\caption{Interpretation of the mutually unbiased bases for left and right qubit}
\vspace{1ex} \centering\footnotesize
\begin{tabular}{c c c c c c c } \hline
& \multicolumn{6}{c}{\textbf{Basis vectors}}
\\[0pt]\cline{1-7}
 Qubit &  $|0\rangle\negthinspace $ & $|1\rangle\negthinspace$ &  $|0'\rangle\negthinspace$ & $|1'\rangle\negthinspace$ & $|0''\rangle\negthinspace$ & $|1''\rangle\negthinspace$
\\[1pt]\hline
\textbf{Left }& (0,1) &  (0,2)& (0,1) & (1,2) &(0,2)  & (1,2)
\\[1pt]\hline
\textbf{Right }& 0 &  1 & 0& 2& 1& 2
\\[1pt]\hline
\end{tabular}\label{tabela222}
\end{table}
Measurement of the left of the entangled qubits (\ref{spl}) changes the state of the right qubit.  By writing the left qubit in the relevant base (which represents the particular type of food given by the offering player, see Table \ref{tabela222}), we are able to determine all possible states of the right qubit after the measurement is made: 
\begin{eqnarray}
&&\nonumber\frac{1}{\sqrt{2}}(|0\rangle\negthinspace\,|0\rangle\negthinspace + |1\rangle\negthinspace\,|1\rangle\negthinspace)=\\\nonumber&&\frac{1}{\sqrt{2}}|0'\rangle\negthinspace\,\otimes\bigg(\frac{|0\rangle\negthinspace + |1\rangle}{\sqrt{2}}\negthinspace\bigg) + \frac{1}{\sqrt{2}}|1'\rangle\negthinspace\,\otimes\bigg(\frac{|0\rangle\negthinspace - |1\rangle}{\sqrt{2}}\negthinspace\bigg)=\\\nonumber&&\frac{1}{\sqrt{2}}|0''\rangle\negthinspace\,\otimes\bigg(\frac{|0\rangle\negthinspace - i |1\rangle}{\sqrt{2}}\negthinspace\bigg) + \frac{1}{\sqrt{2}}|1''\rangle\negthinspace\,\otimes\bigg(\frac{|0\rangle\negthinspace + i |1\rangle}{\sqrt{2}}\negthinspace\bigg)\nonumber
\end{eqnarray}
It should be noticed that with  probability $1/2$, the left qubit reduces itself to one of the vectors of the measured base.
As a result of  $H$ matrix's activity (\ref{sfera}), the above mentioned equations look as follows:       
 \begin{eqnarray}
&&\frac{1}{\sqrt{2}}|0\rangle\negthinspace\, \otimes(a|0\rangle\negthinspace +c|1\rangle\negthinspace\,) +\nonumber\\\nonumber&& \frac{1}{\sqrt{2}}|1\rangle\negthinspace\,\otimes(\frac{b+d}{\sqrt{2}}|0'\rangle\negthinspace\,+\frac{b-d}{\sqrt{2}}|1'\rangle\negthinspace\,)=\nonumber\\\nonumber
\nonumber\\
&&\frac{1}{\sqrt{2}}|0'\rangle\negthinspace\, \otimes(\frac{a+b}{\sqrt{2}}|0\rangle\negthinspace +\frac{c+d}{\sqrt{2}}|1\rangle\negthinspace \,) +\nonumber\\&& \frac{1}{\sqrt{2}}|1'\rangle\negthinspace\,\otimes(\frac{a-b-ic+id}{2}|0''\rangle\negthinspace\,+\frac{a-b+ic-id}{2}|1''\rangle\negthinspace\,)=\nonumber\\
\nonumber\\
&&\frac{1}{\sqrt{2}}|0''\rangle\negthinspace\, \otimes(\frac{a-ib+c-id}{2}|0'\rangle\negthinspace +\frac{a-ib-c+id}{2}|1'\rangle\negthinspace\,) +\nonumber\\\nonumber&& \frac{1}{\sqrt{2}}|1''\rangle\negthinspace\,\otimes(\frac{a+ib-ic+d}{2}|0''\rangle\negthinspace\,+\frac{a+ib+ic-d}{2}|1''\rangle\negthinspace\,)\nonumber
\end{eqnarray}
Squares of the modules of the right qubit's coordinates saved in the relevant base (which corresponds with the proposed pair of food) represent probability of choosing the given type of food from the proposed pair and define mapping  $A_1:D_6\rightarrow S_3$, which determines the relationship between conditional probabilities that we are interested in and a point on sphere  $S_3$ which parameterizes the cat's tactic:
\begin{eqnarray}
\label{warr}
P_0(C_0|B_2)&=|a|^2\,, \qquad &P_0(C_1|B_2)=|b|^2 \,, \nonumber
\\ P_0(C_0|B_1)&=\frac{|b+d|^2}{2}\,, \qquad &P_0(C_2|B_1)= \frac{|b-d|^2}{2}\,,\nonumber
\\P_1(C_0|B_2)&=\frac{|a+b|^2}{2}\,, \qquad &P_1(C_1|B_2)= \frac{|c+d|^2}{2}\,,
\\P_1(C_1|B_0)&=\frac{|a-b-ic+id|^{2}}{4}\,,\qquad &P_1(C_2|B_0)= \frac{|a-b+ic-id|^{2}}{4}\,,\nonumber
\\P_2(C_0|B_1)&=\frac{|a-ib+c-id|^2}{4}\,, \qquad &P_2(C_2|B_1)= \frac{|a-ib-c+id|^2}{4}\,,\nonumber
\\P_2(C_1|B_0)&=\frac{|a+ib-ic+d|^2}{4}\,, \qquad &P_2(C_2|B_0)= \frac{|a+ib+ic-d|^2}{4}.\nonumber
\end{eqnarray}
Using (\ref{sfera}) conditions (\ref{warr}) may be presented in the following manner (for six independent probabilities):
\begin{eqnarray}\label{war2}
 P_0(C_0|B_2)&=&x_1^2+x_2^2\,,   \nonumber
\\P_0(C_0|B_1)&=&\frac{(x_1+x_3)^2 +(-x_2+x_4)^2}{2}\,,  \nonumber
\\P_1(C_0|B_2)&=&\frac{(x_1+x_3)^2 +(x_2+x_4)^2}{2}\,,  
\\P_1(C_1|B_0)&=&\frac{(x_1+x_2-x_3+x_4)^2 +(x_1+x_2+x_3-x_4)^2}{4}\,, \nonumber
\\P_2(C_0|B_1)&=&\frac{(x_1-x_2-x_3+x_4)^2 +(-x_1+x_2-x_3+x_4)^2}{4}\,,  \nonumber
\\P_2(C_1|B_0)&=&x_1^2+x_3^2\,.  \nonumber
\end{eqnarray}
Combination of the $A_1$ projection with $A$ results in the projection $A_2=A\circ A_1$, $A_2: S_3\rightarrow T_2$ of three -- dimensional sphere $S_3$ into a triangle $T_2$. 
\section{Attainability of various type of optimal strategies}\noindent
Here we will present the range of mapping $A_2$ (in comparison with mapping  $A$ -- classical model) illustrating them by 10.000 randomly selected points according to constant distribution of probability on sphere  $S_3$ (Haar measure which maximizes entropy \cite{dun}). We will consider different types of optimal strategies.   
\subsection{Optimal strategies}
\begin{figure}[!ht]
         \centering{
        \includegraphics[
          height=1.9in,
          width=2.1in]%
         {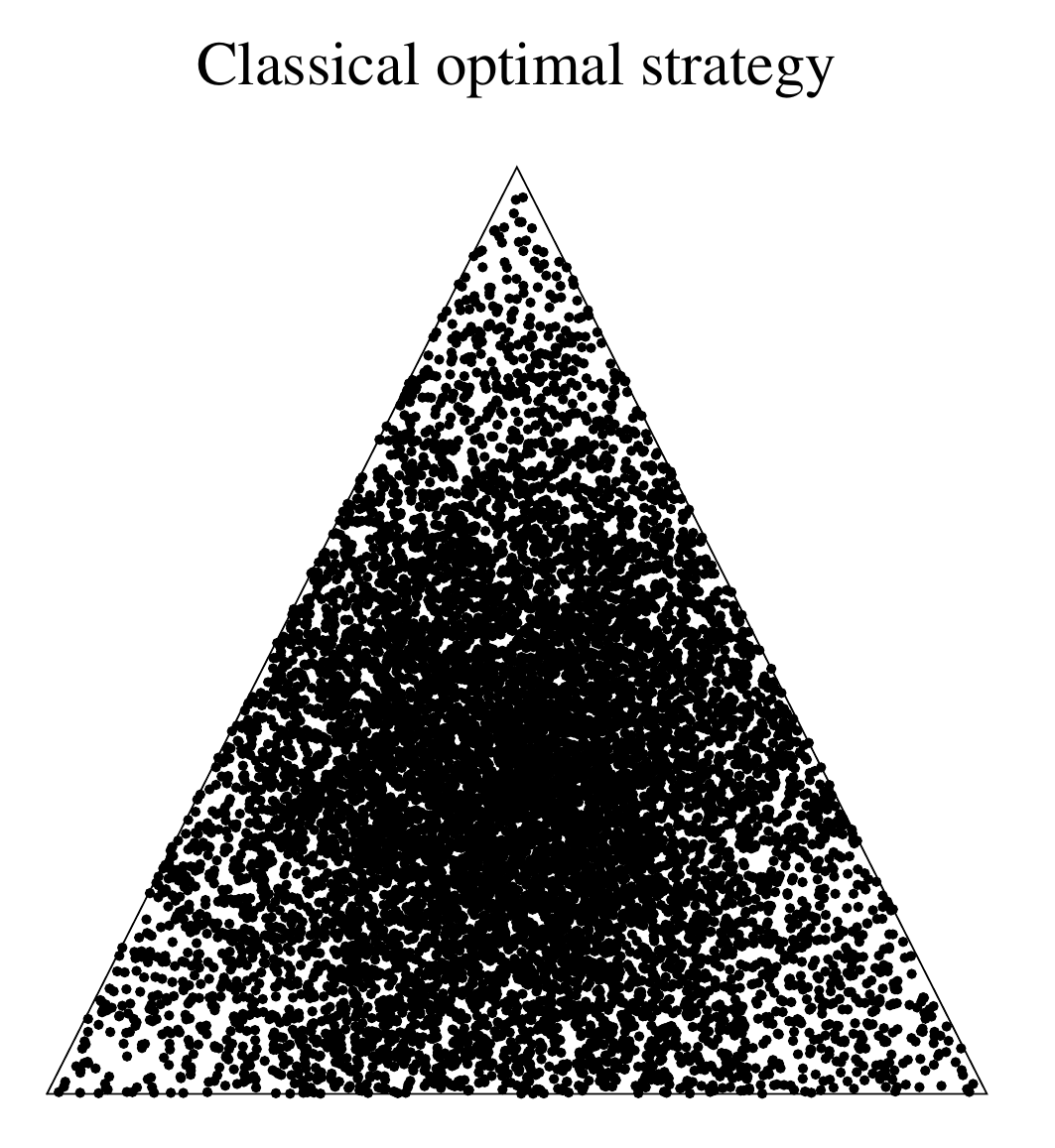}\includegraphics[
          height=1.9in,
          width=2.1in]%
         {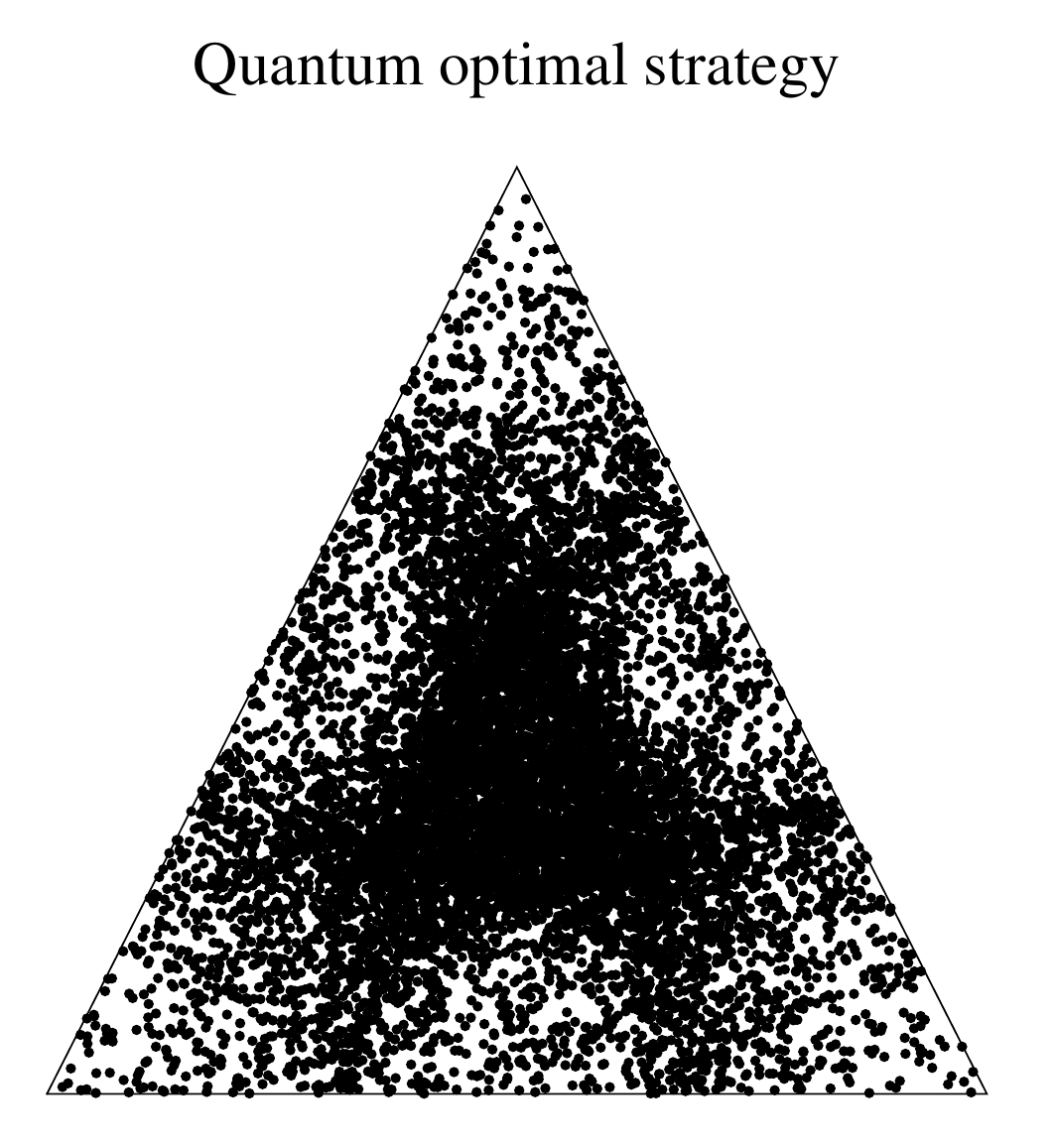}}
         \caption{Optimal strategies.} \label{trt}     
\end{figure}

\noindent
Figure \ref{trt} presents the areas (in both models) of frequencies $(q_0,q_1,q_2)$ for which optimal strategies exist. 
As we can see, the whole area of triangles is covered with dots. This means that in both models (quantum and classical) there exists optimal strategy for  every  triple  $(q_0,q_1,q_2)$. Consider that optimal strategy appears more often in the central part of the triangle, and rarely next to the triangle's vertices where one of the frequencies is larger than the remaining two (one type of food is much more preferred by the offering player than the remaining two).   
\subsection{Intransitive strategies of type 1}
\begin{figure}[!ht]
         \centering{
       \includegraphics[
          height=1.9in,
          width=2.1in]%
         {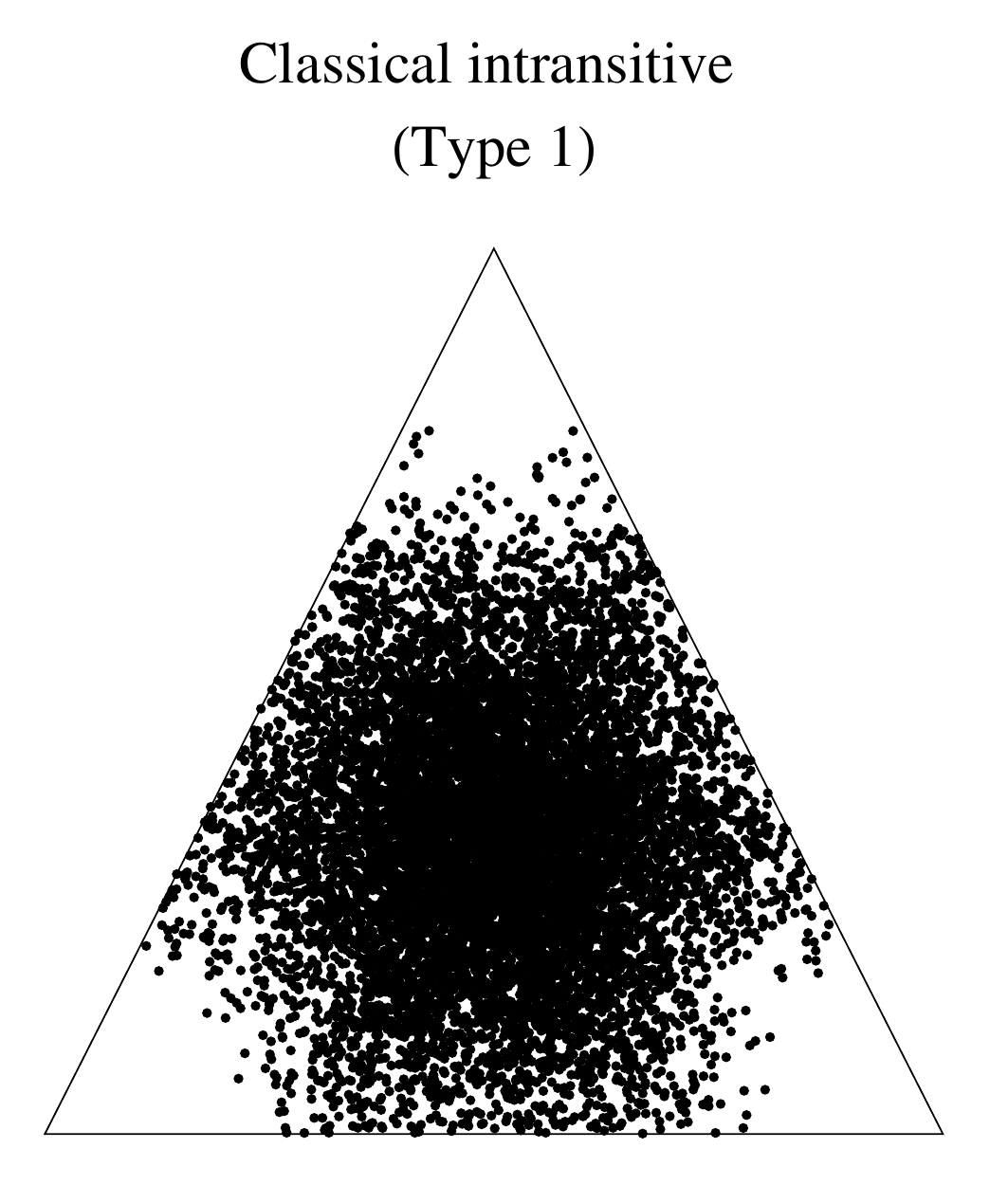}
         \includegraphics[
          height=1.9in,
          width=2.1in]%
         {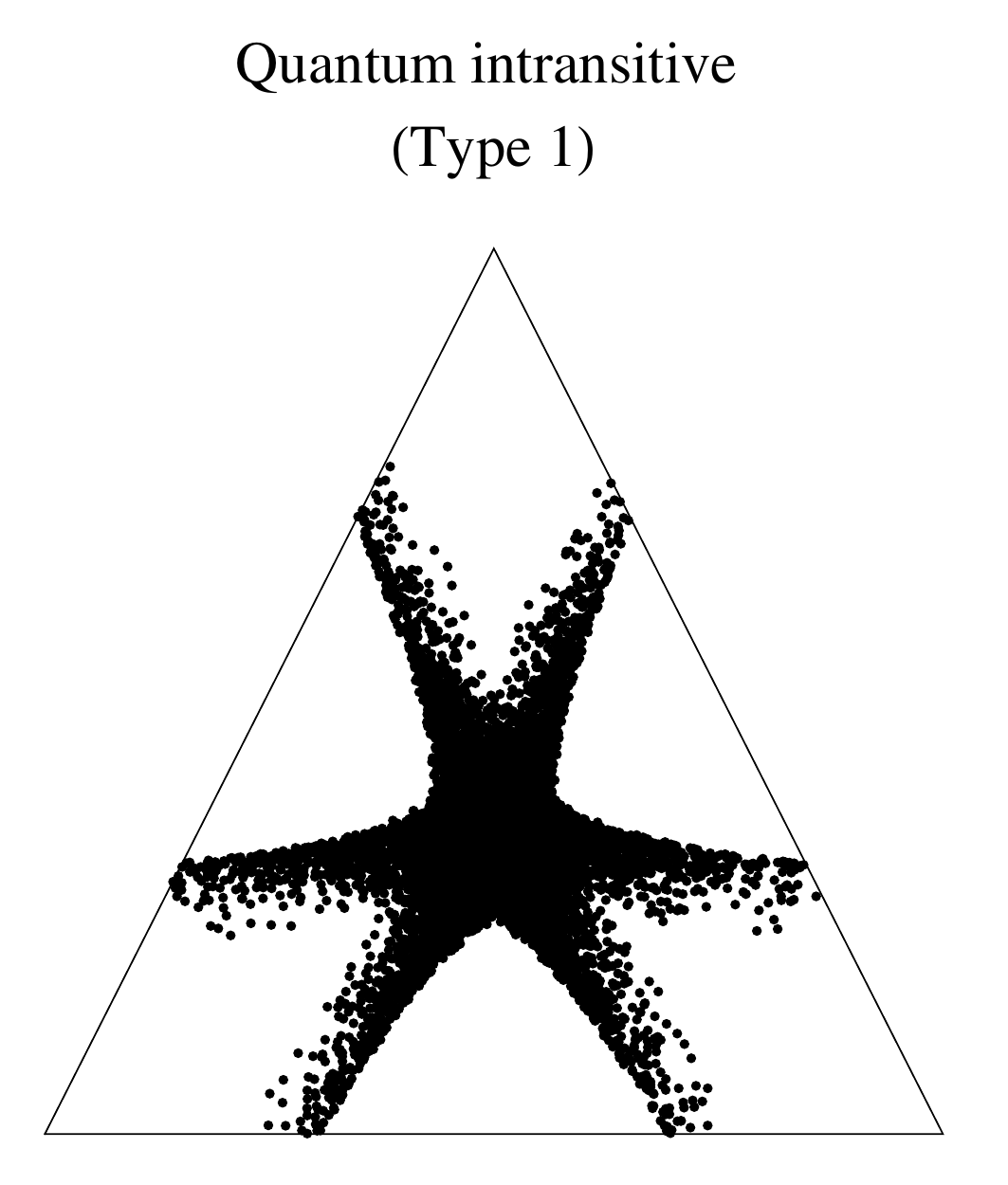}}
         \caption{Optimal intransitive strategies of type 1.}
         \label{3}
\end{figure}

\noindent
Remember that the move of the offering player consists in proposing one out of two pairs containing this type of food (with equal probability). Therefore, the selecting player (the cat) may be forced to choose food from the same pair in two different contexts. Pair $(0,1)$ may be for instance offered when the offering player decides to give type of food no. $0$ or type of food no. $1$. This modification of the game model requires us to specify relation of preferences $\succ$ between the given types of food. We will say that: 
 \begin{equation}\label{rel}
 \emph{food no ~0} \prec \emph{food no ~1}\,,
\end{equation}
if the cat will be always more willing to choose type of food $1$ rather than type of food $0$ from pair $(0,1)$ ($P_0(C_1|B_2)>P_0(C_0|B_2)$ and $P_1(C_1|B_2)>P_1(C_0|B_2)$). We deal with an intransitive choice (in above discussed sense) if  one of the following conditions is fulfilled:
\begin{enumerate}
	\item 
 \begin{eqnarray}\label{Typ11}
P_0(C_0|B_2)<\frac{1}{2}\,, &\qquad & P_1(C_0|B_2)<\frac{1}{2} \,,\nonumber 
\\P_1(C_1|B_0)<\frac{1}{2}\,,& \qquad & P_2(C_1|B_0)<\frac{1}{2} \,, 
\\P_0(C_0|B_1)>\frac{1}{2}\,, &\qquad & P_2(C_0|B_1)>\frac{1}{2} \,.\nonumber
\end{eqnarray}
	\item 
	\begin{eqnarray}\label{Typ12}
P_0(C_0|B_2)>\frac{1}{2}\,, &\qquad& P_1(C_0|B_2)>\frac{1}{2} \,, \nonumber
\\P_1(C_1|B_0)>\frac{1}{2}\,,& \qquad& P_2(C_1|B_0)>\frac{1}{2} \,, 
\\P_0(C_0|B_1)<\frac{1}{2}\,,& \qquad& P_2(C_0|B_1)<\frac{1}{2} \,.\nonumber
\end{eqnarray}\end{enumerate}
Figure \ref{3} presents the areas (in both models) of frequencies $(q_0,q_1,q_2)$ for which optimal intransitive strategies exist (type 1 -- conditions (\ref{Typ11}) or (\ref{Typ12})).
There is a significant difference in the way of achieving this type of strategies in both models. In classical model, optimal intransitive strategy does not occur for frequencies represented by points of the triangle that are located next to its vertices (when one of the frequencies is significantly higher than others). On the other hand, in quantum model this area is much more restricted and focuses in the central part of the triangle. 
  
\subsection{Intransitive strategies of type 2}
\begin{figure}[!ht]
         \centering{\includegraphics[
          height=1.9in,
          width=2.1in]%
         {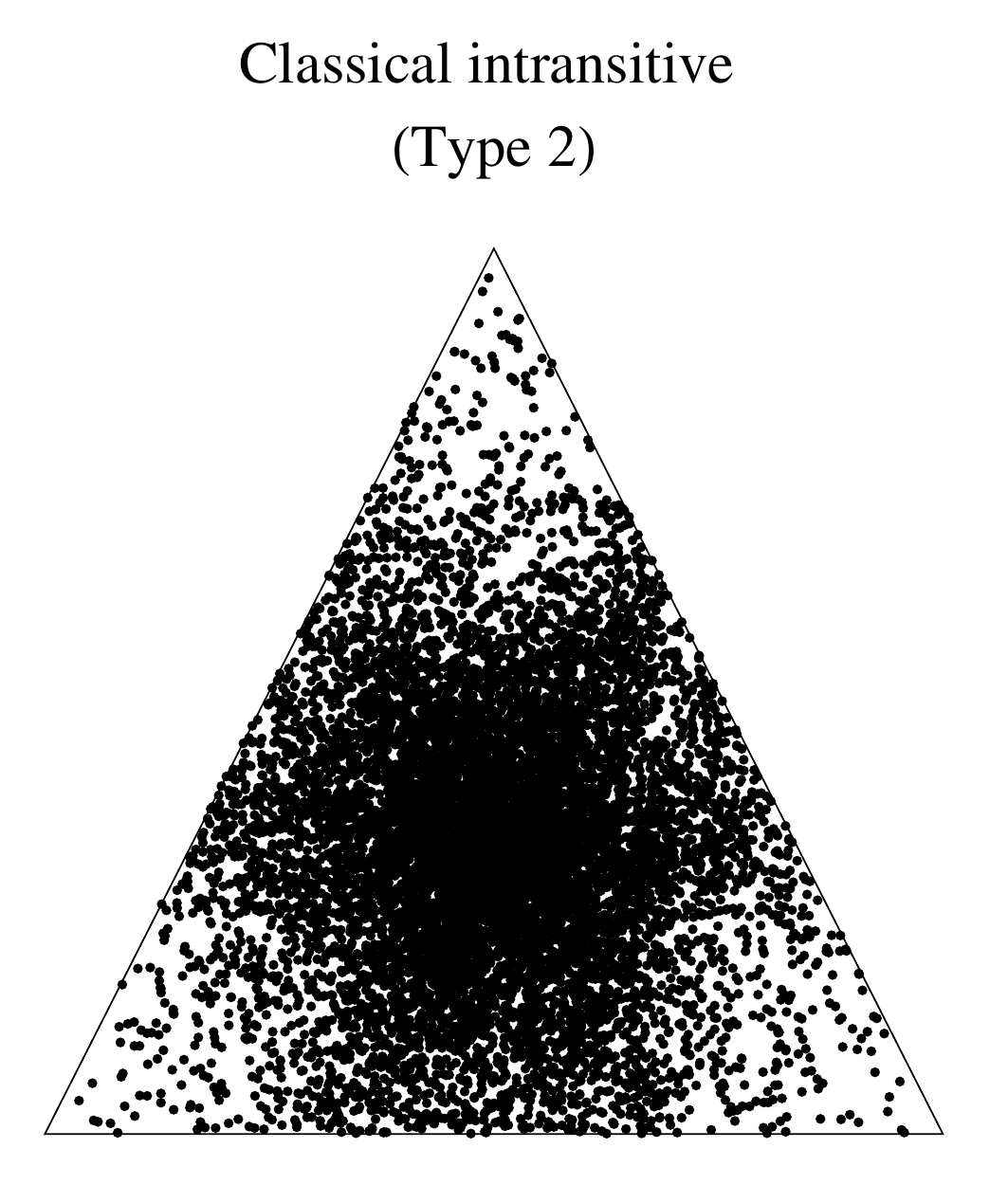}
       \includegraphics[
          height=1.9in,
          width=2.1in]%
         {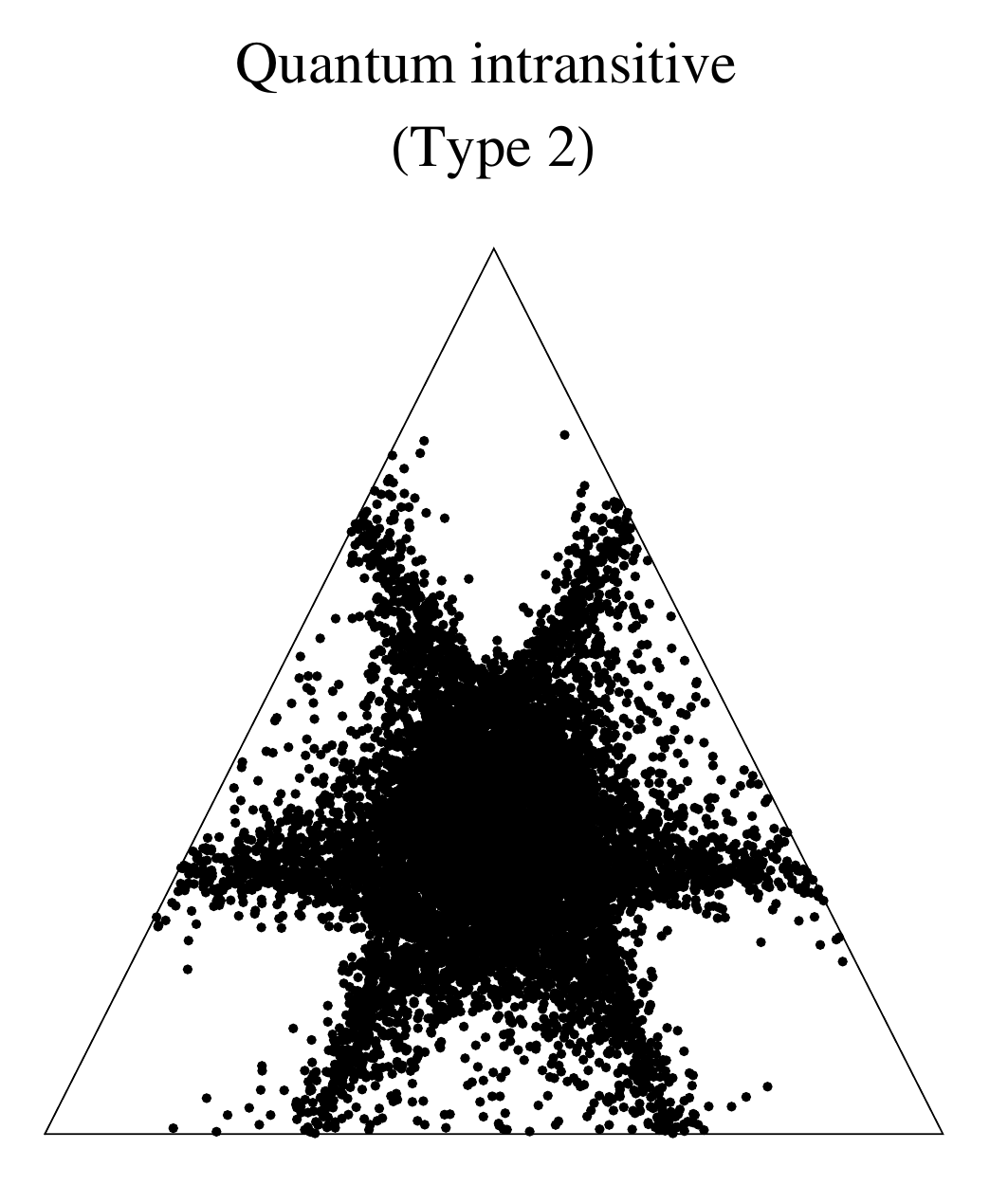}}\caption{Optimal intransitive strategies of type 2.}\label{4}
\end{figure}

\begin{figure}[!ht]
         \centering{
       \includegraphics[
          height=1.9in,
          width=2.1in]%
         {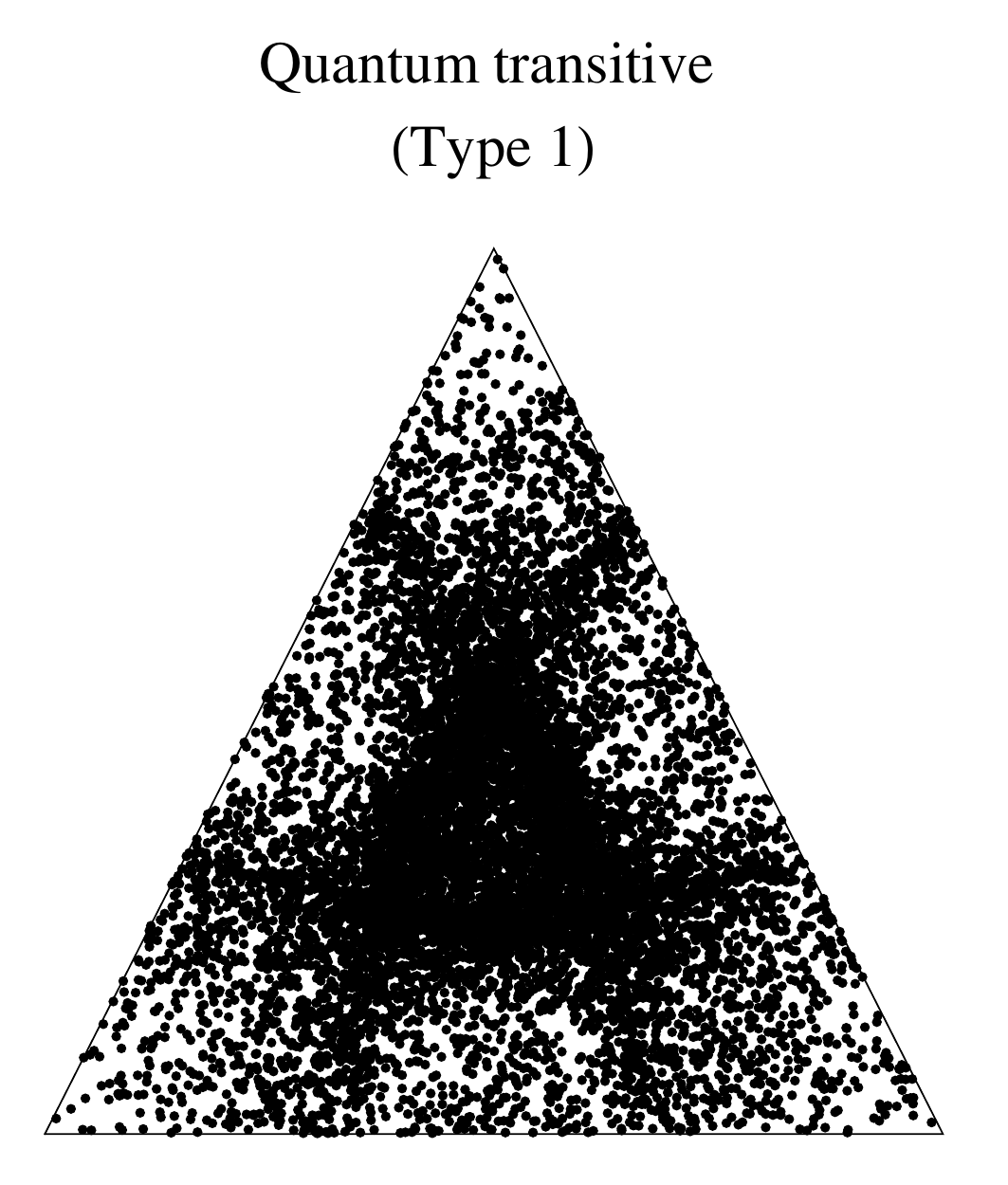}
         \includegraphics[
          height=1.9in,
          width=2.1in]%
         {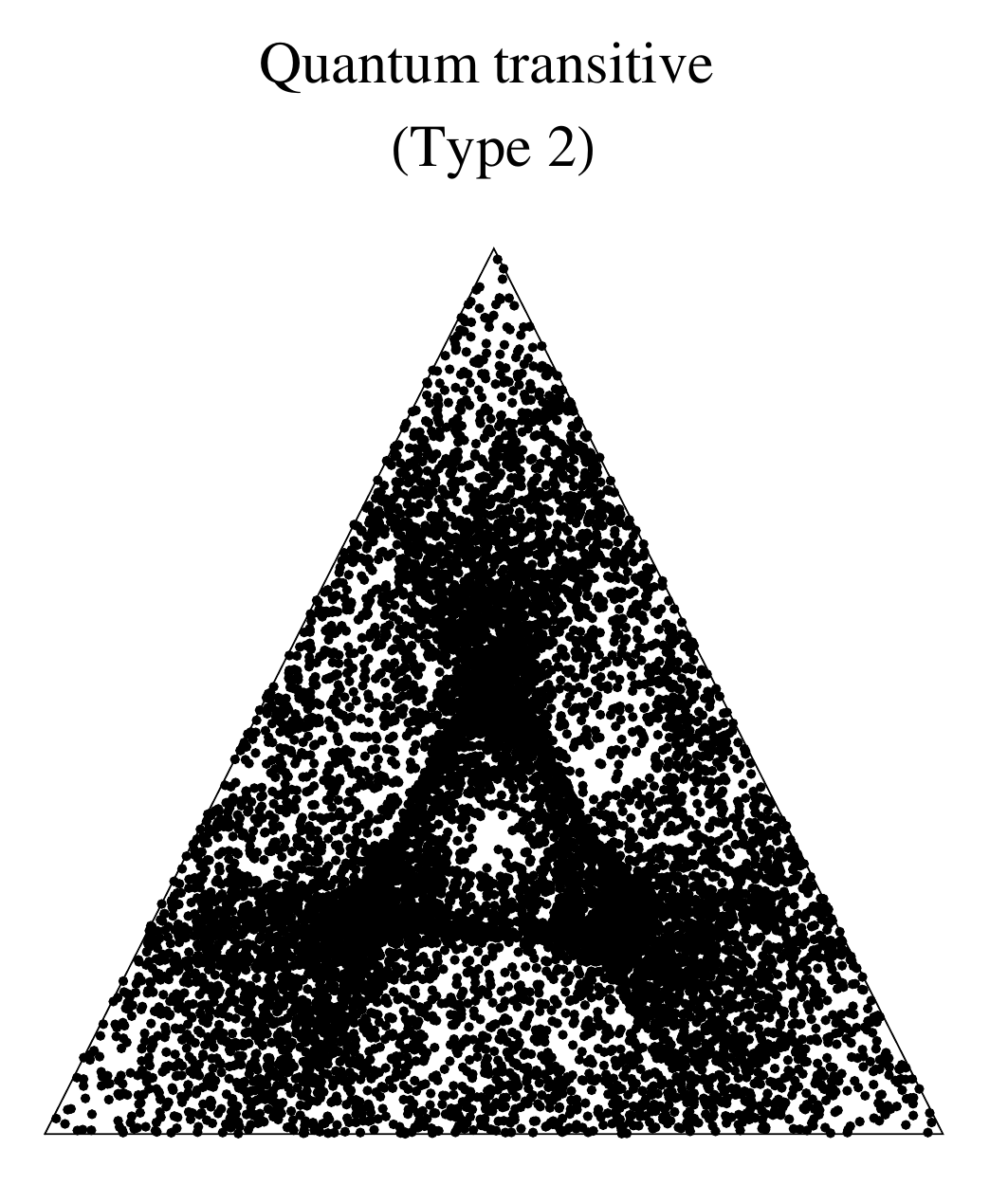}}
         \caption{Optimal transitive strategies.}
         \label{sint}
\end{figure}

\noindent
Relation of preferences between the types of food may also be  defined differently. The cat prefers the given food $x$ more than food $y$ when total frequency of choosing food $x$ out of pair $(x,y)$ which appears in two different contexts, is larger than total frequency of choosing food $y$. 
The cat prefers food no. $1$ to food no. $0$ (\ref{rel}) if the following condition is satisfied:
\begin{displaymath}
P_0(C_1|B_2)+P_1(C_1|B_2)>P_0(C_0|B_2)+P_1(C_0|B_2)\,.
\end{displaymath}
Assuming such definition of preferences relation, conditions for the strategies' intransitivity will look as follows:  
\begin{enumerate}
\item 
 \begin{eqnarray}\label{Typ21}
P_0(C_0|B_2)+P_1(C_0|B_2)&<1\,, \nonumber
\\P_1(C_1|B_0)+P_2(C_1|B_0)&<1\,, 
\\P_0(C_0|B_1)+P_2(C_0|B_1)&>1\,,\nonumber
\end{eqnarray}
\item 
\begin{eqnarray}\label{Typ22}
P_0(C_0|B_2)+P_1(C_0|B_2)&>1\,, \nonumber
\\P_1(C_1|B_0)+P_2(C_1|B_0)&>1\,, 
\\P_0(C_0|B_1)+P_2(C_0|B_1)&<1\,.\nonumber
\end{eqnarray}\end{enumerate}
Figure \ref{4} presents the areas (in both models) of frequencies $(q_0,q_1,q_2)$ for which optimal intransitive strategies (type 2--conditions (\ref{Typ21}) or (\ref{Typ22})) exist.
We can observe that in the classical model for any frequencies  $(q_0, q_1, q_2)$, there always exists an optimal intransitive strategy. In quantum model, such strategies does not occur for frequencies $(q_0, q_1, q_2)$ located in the vicinity of the triangle's vertices. We are able to observe distinct concentration of optimal strategies on the area of the six--arm star (on other areas they are less numerous). 
\vspace{0.2cm}\\
Different definition intransitive orders, yielded remarkably distinct results. This is a direct consequence of conditions influencing the type of given intransitivity. In the first case (type 1) relation of order depends on much stronger conditions than in the second case (type 2). The selecting player who prefers food $x$ to food $y$ should more often choose type of food $x$ from pair $(x, y)$ in two possible situations depending on the context. The intransitivity of type 2 is based on much weaker relation of order, according to which total frequency of choosing food $x$ in two different situations when pair $(x, y)$ appears should be larger than total frequency of choosing food $y$ from the same pair. Intransitivity of type 1 is, therefore, only a special case the more general type 2. 
\vspace{0.2cm}\\
Let us look at areas of the triangle which correspond with transitive strategies in quantum model (Figure \ref{sint}). 
It turns out that in case of both relations of preferences, transitive optimal strategies exists for all points of the triangle. It implies that for all triple
$(q_0,q_1,q_2)$, if there exists an optimal intransitive strategy, there must exist corresponding transitive strategy as well.
\section{Conclusions} \noindent
In this article we carried out an analysis of the simple choice game by changing the construction of the game known from previous analyses \cite{r2,rMM,r1}.  The proposed modification eliminated the limitations which made it impossible (in some cases) for the selecting player to achieve optimal strategies. It turns out that in the new conditions optimal intransitive strategies in the quantum model lose their importance (because all strategies of this type may be replaced by transitive strategies having the same , i.e. optimal, effect). Moreover, the areas of simplex $T_2$ corresponding with the intransitive strategies (type 1 and type 2) are smaller in the quantum model than in the classical model. Therefore, intransitive strategies are represented more rarely in the quantum model. In the previous game version, intransitive strategies have significant importance in the quantum model (there are frequencies $(q_0,q_1,q_2)$ for which the optimum effect may be obtained only by intransitive strategy). This model was based on assumptions limiting the existence of an optimal strategy. It is not surprising that introducing an ``automatic machine'' (which distributed the given food pairs with equal probability) to the game increased the attainability of optimal strategies (they exist for any point of triangle $T_2$). Indeed, the fact that this modification caused loss of importance of the intransitive strategies in the quantum model seems to be much more interesting. 
Perhaps those players (people) who are subject to a large number of limitations make ``intransitive decisions'' more often, while those who are presented with a wider range of options tend to make transitive choices. In this article we only presented a simple game model. It is still open to discussion whether the characteristics observed in this example will prove to be valid in more complicated and sophisticated models and whether they constitute an authentic model of behaviours.        
\vspace{0.2cm}\\
In games, the players' moves are in a sense interconnected, i.e. one player's move urges the second player to react. The quantum model of the game presented in this article was based on a pair of qubits in an entangled state (where the qubits' features are correlated). The entangled state seems to be a natural tool for modelling game interactions. In many games, the quantum entangled state led to surprising conclusions and observations \cite{r9}. Therefore, it is worthwhile to take a closer look at the importance of intransitive strategies in the context of quantum entangled states.   


\end{document}